\documentclass[12pt]{iopart}
\usepackage{dcolumn}% Align table columns on decimal point
\usepackage{bm}% bold math
\usepackage{graphicx}% Include figure files
\usepackage{color}
\usepackage{subfigure}
\usepackage{hyperref}
\usepackage{latexsym}
\usepackage{amsthm}
\usepackage{amssymb}
\usepackage{cite}
\DeclareGraphicsExtensions{.jpg,.pdf, .mps, .png, .eps, .ps, .EPS,.gif}

\begin{document}
\def\be{\begin{equation}}
\def\ee{\end{equation}}

\def\bc{\begin{center}}
\def\ec{\end{center}}
\def\bea{\begin{eqnarray}}
\def\eea{\end{eqnarray}}
\newcommand{\avg}[1]{\langle{#1}\rangle}
\newcommand{\Avg}[1]{\left\langle{#1}\right\rangle}

\def\ie{\textit{i.e.}}
\def\etal{\textit{et al.}}
\def\m{\vec{m}}
\def\G{\mathcal{G}}

\title[Large deviation theory of percolation on multiplex networks]{Large deviation theory of percolation on interdependent multiplex networks }

\author{Ginestra Bianconi}

\address{Alan Turing Institute, London, United Kingdom\\
School of Mathematical Sciences, Queen Mary University of London, London, United Kingdom}
\ead{g.bianconi@qmul.ac.uk}
\vspace{10pt}
\begin{indented}
\item[]
\end{indented}

\begin{abstract}
Recently increasing attention has been addressed to the fluctuations observed in percolation defined in single and multiplex networks. These fluctuations are extremely important to characterize the robustness of real finite networks but cannot be captured by the traditionally adopted mean-field theory of percolation.
Here we propose a theoretical framework and  a message passing algorithm that is able to fully capture the large deviation of percolation in interdependent multiplex networks  with a locally tree-like structure.
This framework is here applied to study the robustness of single instance multiplex networks and compared to the results obtained using extensive simulations of the initial damage.
For simplicity the method is here  developed for interdependent multiplex networks without link overlap, however it can  be  generalized to treat multiplex networks with link overlap.
\end{abstract}

\section{Introduction}

Percolation is a fundamental critical phenomena \cite{crit,Kahng_review} defined in complex networks as it sheads light on the robustness of networks when a fraction $f=1-p$ of nodes is randomly removed. In the last ten years important new insights on percolation theory have been gained by considering percolation processes defined on multiplex networks \cite{Havlin,Baxter,Son,bianconi2018multilayer,Havlin2,Havlin3,Cellai1,Cellai2,Baxter2,Goh_comment,Radicchi_Bianconi,Goh1,Kahng,Ivan}.  
Multiplex networks \cite{bianconi2018multilayer,PhysicsReports,Kivela,Goh_review,Bianconi2013} describe generalized network structures formed by several interacting networks (layers). Examples of multiplex networks include global  infrastructures, interacting financial networks and  biological networks in the cell or in the  brain. 

Multiplex networks are often characterized by interdependencies \cite{Havlin} existing between nodes of different layers implying that the failure of one node in one layer necessarily causes the failure of the interdependent nodes in the other layers independently of the rest of the network.
Interestingly interdependent multiplex networks are much more fragile than single networks \cite{Havlin,Baxter,Son,bianconi2018multilayer}. In particular  the percolation threshold $p=p_c$ of multiplex networks is larger than the percolation threshold of their single layers taken in isolation. Moreover if the layers are not formed by identical networks,  the percolation transition is discontinuous and  hybrid and  characterized by large avalanches of failures going back and forth among different layers. These large avalanches of failure events  are the most notable effect of interdependencies and they are a strong signal of the increased fragility of multiplex networks.

The order parameter of the percolation defined on interdependent multiplex networks is given by the Mutually Connected Giant Component (MCGC) \cite{Havlin,Baxter} which generalizes the giant component and is formed by the set of nodes such that each pair of nodes is connected by a path in at least one layer.
In infinite interdependent multiplex networks the percolation transition can be studied using a suitable mean-field approach to percolation\cite{Havlin,Baxter,Son,bianconi2018multilayer}. In fact in multiplex network as well as in single networks, percolation theory is self-averaging and in the infinite network limit the mean-field approach gives exact results.
In finite  interdependent multiplex networks when the initial damage configuration is known, the percolation transition on interdependent multiplex networks can be described by a message passing algorithm \cite{Son,bianconi2018multilayer,Bianconi_Radicchi,Radicchi} that is able to predict which nodes are in the MCGC after the initial damage provided the multiplex network is locally tree-like.  Interestingly this message passing algorithm has a definition that depends  whether the multiplex network displays or not link overlap, i.e. it depends wheather there are  pairs of nodes connected in more than one layer \cite{Cellai1,Cellai2,Bianconi_Radicchi,Baxter2,Goh_comment}.
When the specific nature of the initial damage configuration is not known the  message passing algorithm can be averaged over the distribution of the initial damage. This mean-field approach provides the probability that each node is in the MCGC when the initial damage configuration is drawn from the known distribution (for instance when each node is damaged with probability $f=1-p$). However this approach is not able to capture the fluctuations that can be observed in the size of the MCGC for different realizations of the initial damage.

Recently, there has been a surge of interest in characterizing the risk of dramatic failure of events as a consequence of a random damage \cite{Fluctuations,rare,Radicchi3,Krapivsky,Caldarelli,other,Kahng2}. In fact real networks are typically finite and often sufficiently far from the thermodynamic limit. Therefore   the outcome of a initial damage configuration drawn from a given distribution  can have large fluctuations not predicted by the mean-field approach to percolation.  Characterizing these fluctuations is of fundamental importance for applications as the average response to perturbation captured by the mean-field theory of percolation can be dramatically misleading \cite{Radicchi3,other}. Moreover characterizing the node sets that are responsible for safeguarding the cohesiveness of a network \cite{Radicchi3} or for more efficiently disrupting a network (optimal percolation) \cite{Makse,Lenka,Osat,Mendes} have a number of applications. These range from the design principles for robust infrastructures to the control of epidemic processes on networks.
From the theoretical perspective characterizing these fluctuations is very interesting as well as  it sheds new light on the nature of the percolation transition \cite{Fluctuations,rare,Krapivsky,Caldarelli,Kahng2}.

These phenomena are naturally addressed using the  theory of large deviations \cite{Touchette, Hollander}.
On single networks the characterization of the large deviation properties of random graphs  has been first tackled using the properties of the Ising model \cite{ChayesL} and the Potts model \cite{Monasson,Bradde}. While the study of the Potts model sheads light on the large deviations of the number of components \cite{Monasson,Bradde} the properties of the Ising model can characterize the large deviations of the giant component in random Erd\"os and R\'enyi graphs \cite{ChayesL} with constant average degree. However the analytical method developed in Ref. \cite{ChayesL} is not able to predict the distribution of the giant component in a real networks or in a network model different from a Erd\"os and R\'enyi graph. 
In order to address this problem a sofisticated numerical technique based on a Markov Chain Monte Carlo method has been shown \cite{Hartmann1} to be able to uncover the full distribution of the giant component down to rare events with probabilities $10^{-100}$. This method has subsequently been also used  for numerically evaluating the distribution of the diameter \cite{Hartmann_Mezard} and of the largest biconnected component of networks\cite{Hartmann2}.

Recently \cite{rare} a large deviation theory of percolation has been proposed to capture the fluctuations observed in percolation of   single finite  networks. This approach is based on message-passing algorithm and specifically on Belief Propagation \cite{Mezard_Montanari} and can be applied to single instances of networks provided that they are locally tree-like. This work reveals that when one considers aggravating initial damage configurations, the percolation transition can become discontinuous also in percolation defined on single networks.

Here we generalize the large deviation theory of percolation to multiplex networks. For simplicity we consider  only multiplex networks without link overlap. However it is possible to  generalize the method  to multiplex networks with link overlap.  We show that the percolation transition remains  discontinuous for both aggravating and buffering configuration of the initial damage. Interestingly we can measure the fluctuations in the size of the MCGC  and we observe  that these fluctuations can remain very significant up to the percolation transition in the typical scenario. Finally we use the large deviation theory to theoretically predict the convex envelop of the rate function $I(R)$ of the distribution $\pi(R)$ the size $R$ of the   MCGC   finding very good agreement with extensive numerical simulations.

The paper is structured as follows. In Sec. 2 we present the mean-field theory of percolation of interdependent duplex networks without link overlap.
In Sec. 3 we present  the large deviation theory of percolation. In Sec. 4 we show how the Belief Propagation approach can be used to fully characterize the large deviation theory of percolation.
In Sec. 5 we compare the numerical results obtained with the Belief Propagation approach with extensive numerical simulations. Finally in Sec. 6 we give the conclusions.

\section{The mean-field theory of percolation in interdependent multiplex networks}
\subsection{Multiplex networks and mutually connected giant component}

We consider  an interdependent duplex network $\vec{G}=(G^{[1]},G^{[2]})$ formed by two networks $G^{[\alpha]}=(V,E^{[\alpha]})$ (with $\alpha=1,2$) among the same set of $N$ nodes $V=\{i|i\in\{1,2,\ldots, N\}\}$ \cite{Bianconi2013,bianconi2018multilayer}. Each network $\alpha$ is defined by the adjacency matrix ${\bf a}^{[\alpha]}$.
For each node $i$ we consider two replica nodes $(i,1)$ and $(i,2)$ indicating the identity of node $i$ in layer 1 and in layer 2 respectively. Each pair of replica nodes is connected by an {\it interlink} that has a different valence with respect to the links present in each layer. In fact in interdependent multiplex networks interlinks indicate interdependencies whose role in determining the robustness properties of the duplex network is explained in the following.
Duplex networks can be classified depending on the presence or the absence of link overlap. The total  overlap between the two layers of a duplex network is given by the number of pairs of nodes that are connected in both layers, i.e.
\bea
O=\sum_{i<j} a_{ij}^{[1]}a_{ij}^{[2]}.
\eea
 We therefore call duplex networks without link overlap the duplex networks in which the total overlap vanishes and we call duplex networks with link overlap the duplex networks in which the total overlap is greater than zero. 

We monitor the robustness of the considered duplex networks by studying the size of  the Mutually Connected Giant Component (MCGC) \cite{Havlin} after an initial damage of a fraction $f=1-p$ of the nodes. Due to the presence of interdependencies, replica nodes  cannot be in the MCGC if their corresponding replica node is not included in it. Therefore the MCGC can be evaluated by propagating the initial failure back and forth among the different layers of the duplex network \cite{Havlin,bianconi2018multilayer}.
Specifically the algorithm that define the MCGC  prescribes to first evaluate the giant component in each individual layer  and subsequently to damage each replica node whose corresponding  interdependent replica node is not in the giant component. This algorithm is then iterated until the damage does not propagate any more among different layers.  
Alternatively the  MCGC can be  defined as the giant component formed by pairs of nodes connected to each other at least by a path in each layer \cite{Baxter}. 

\subsection{Message passing algorithm on single network and single realization of damage}
On a  locally tree-like interdependent duplex network without link overlap percolation can be described by the following message passing equations.
We assume that initially each node of the duplex network is damaged randomly with probability $f=1-p$, and we associated a variable $x_i=1$ to a node $i$ that is not initially damaged and a variable $x_i=0$ to  a node $i$ that is initially damaged. 
Additionally we assign to each node $i$ a variable $\rho_i$ indicating if the node belongs to the MCGC $(\rho_i=1)$ or not $(\rho_i=0)$. 
The indicator function $\rho_i$ depends on the configuration of the initial damage $\bm{x}=\{x_i|i\in\{1,2,\ldots, N\}\}$ and  the messages $\sigma_{i\to j}^{\alpha}$ that each node $i$ send to a neighbour node $j$ in layer $\alpha$. 
In particular a node $i$ belongs to the MCGC if it is not initially damaged and receives at least one positive message in each layer,i.e.
\bea
\rho_i=x_i\prod_{\alpha=1}^2\left[1-\prod_{\ell\in N_{\alpha}(i)}(1-\sigma_{\ell\to i}^{\alpha})\right].
\label{rho0}
\eea  
The messages $\sigma_{i\to j}^{\alpha}\in\{0,1\}$ take the value $\sigma_{i\to j}^{\alpha}=1$ if the node $i$ is not initially damaged and if in each layer node $i$ receives at least a  positive message  from a neighbour nodes different from node $j$, i.e.
\bea
\sigma_{i\to j}^{\alpha}=x_i\left[1-\prod_{\ell\in N_{\alpha}(i)\setminus j}(1-\sigma_{\ell\to i}^{\alpha})\right]
\left[1-\prod_{\ell\in N_{\beta}(i)}(1-\sigma_{\ell\to i}^{\beta})\right],\label{mes0}\eea
where here and in the following $\beta$ indicates the layer $\beta\in\{1,2\}$ with $\beta\neq \alpha$.
The  size ${\mathcal R}$ of the MCGC is given by the number of nodes that belong to it, i.e.
\bea
{\mathcal{R}}=\sum_{i=1}^N \rho_i.
\eea
We note that different initial damages can yield MCGC of different sizes. The goal of this paper  is to uncover the fluctuations that can be observed in the sizes of the MCGC resulting from comparable configurations of the initial damage. \\

We stress here that Eqs. (\ref{rho0}),(\ref{mes0}) are only valid in absence of link overlap. In fact when the multiplex network display link overlap these equations only define the Directed Mutually Connected Component (DMCGC) which can be considered as the outcome of an epidemic spreading dynamics in which each node can become infected only if is in contact with  at least an infected individual in each layer \cite{Cellai1,Cellai2}.
In presence of link overalp the MCGC is instead captured by a message passing algorithm in which the messages are not scalars like $\sigma_{i\to j}^{\alpha}$ but they are instead $M$-dimensional vectors $\vec{n}_{i\to j}$ of elements $n_{i\to j}^{\alpha}\in \{0,1\}$. Here $n_{i\to j}^{\alpha}$ indicates wherether ($n_{i\to j}^{\alpha}=1$)  or not ($n_{i\to j}^{\alpha}=0$) node $i$ connects node $j$ to the MCGC through layer $\alpha$ as long as node $j$ is assumed to be in the MCGC \cite{Cellai2,Baxter2}.
These message elements are set of one ($n_{i\to j}^{\alpha}=1$) if the following two conditions are met:
\begin{itemize}
\item[(i)]
node $i$ is in the MCGC;
\item[(ii)] node $i$ is connected in layer $\alpha$ to at least one node of the MCGC different from node $j$. 
\end{itemize}
Additionally  node $i$ is in the MCGC if  for any layer of the multiplex network node $i$ is either connected to node $j$ or connected to the MCGC through nodes different from $j$. 
Therefore the message passing equations for $n_{i\to j}^{\alpha}$ read
\bea
n_{i\to j}^{\alpha}=x_i a_{ij}^{[\alpha]}\delta(v_{i\to j},M)\left[1-\prod_{\ell\in N_{\alpha}(i)\setminus j}\left(1-n_{\ell\to i}^{\alpha}\right)\right],
\label{mess_ov}
\eea
where $\delta(c,d)$ indicates the Kronecker delta and $v_{i\to j}$ is defined as
\bea
v_{i\to j}=\sum_{\alpha=1}^M\left[1-\left(1-a_{ij}^{[\alpha]}\right)\prod_{\ell\in N_{\alpha}(i)\setminus j}\left(1-n_{\ell\to i}^{\alpha}\right)\right].
\label{v_ov}
\eea
Therefore the message elements $n_{i\to j}^{\alpha}$ are correlated and this property of these message passing equations considerably complicates the algorithm.
Finally in presence of link overlap the indicator function $\sigma_i$ determining if a node is in the MCGC satisfies 
\bea
\sigma_i=x_i\prod_{\alpha=1}^M\left[1-\prod_{\ell\in N_{\alpha}(i)}\left(1-n_{\ell\to i}^{\alpha}\right)\right].
\label{rho_ov}
\eea
Naturally a close inspection to the message passing equations in presence of link overlap show that these equations reduce to the Eqs. (\ref{rho0}) and (\ref{mes0}) in absence of link overlap. Moreover we note that in absence of link overlap the equations for the MCGC and for the DMCGC coincide.

We note that the message passing Eqs. $(\ref{mess_ov}),(\ref{v_ov})$ and $(\ref{rho_ov})$ can be also written in terms of the indicator functions $\sigma_{i\to j}^{\vec{n}} \in \{0,1\}$ that indicates if node $i$ sends to node $j$ the message $\vec{n}$ $(\sigma_{i\to j}^{\vec{n}}=1)$ or not ($\sigma_{i\to j}^{\vec{n}}=0$). For a detailed account of the message passing theory for percolation of interdependent multiplex networks see Ref. \cite{Cellai2,bianconi2018multilayer}.\\

Here and in the following we will consider exclusively multiplex network without link overlap.
In the following we will indicate with $\bm{\sigma}$ the set of all the messages and with $\bm{\sigma}_i$ the set of all the messages starting or ending at node $i$, i.e.
\bea
\bm{\sigma}&=&\{\sigma_{i\to j}^{\alpha}\}_{\alpha=1,2;i\in\{1,2,\ldots, N\}; j\in N_{\alpha}(i) },\nonumber \\
\bm{\sigma}_i&=&\{\sigma_{i\to j}^{\alpha},\sigma_{j\to i}^{\alpha}\}_{\alpha=1,2; j\in N_{\alpha}(i)}.
\eea
Additionally we will indicate with $\bm{x}$  the configuration of the initial damage, i.e.
\bea
\bm{x}=\{x_i\}_{i\in \{1,2,\ldots, N\}}.
\eea

\subsection{Random realization of the damage and typical behaviour}

Let us consider  initial damage configurations $\bm{x}$ obtained by  damaging each node with probability $1-p$, i.e. each configuration $\bm{x}$ is drawn from a distribution
\bea
\tilde{P}(\bm{x})=\prod_{i=1}^N p^{x_i}(1-p)^{1-x_i}.
\label{random}
\eea
In the mean-field theory of percolation the expected size  $\hat{R}$ of the MCGC given by 
\bea
\hat{R}=\sum_{\bm{x}}\tilde{P}(\bm{x}) {\mathcal R}, 
\eea
is obtained by averaging the  original message passing algorithm  over the distribution $\tilde{P}(\bm{x})$.
On a locally tree-like multiplex network without link overlap this procedure reduces to considering a message passing algorithm 
\bea
\hat{\sigma}_{i\to j}^{\alpha}=p\left[1-\prod_{\ell\in N_{\alpha}(i)\setminus j}(1-\hat{\sigma}_{\ell\to i}^{\alpha})\right] \left[1-\prod_{\ell\in N_{\beta}(i)}(1-\hat{\sigma}_{\ell\to i}^{\beta})\right],
\label{mess_av}
\eea 
where the messages $\hat{\sigma}_{i\to j}^{\alpha}$  correspond to the average of   the messages $\sigma_{i\to j}^{\alpha}$  over the distribution $\tilde{P}(\bm{x})$, i.e.
\bea
\hat{\sigma}_{i\to j}^{\alpha}=\sum_{\bm{x}}\tilde{P}(\bm{x}) \sigma_{i\to j}^{\alpha}.
\eea
Similarly the probability $\hat{\rho}_i$ that node $i$ is in the MCGC
can be obtained starting from the messages $\hat{\sigma}_{i\to j}^{\alpha}$ as 
\bea
\hat{\rho_i}=p\prod_{\alpha=1,2}\left[1-\prod_{\ell\in N_{\alpha}(i)}(1-\hat{\sigma}_{\ell\to i}^{\alpha})\right]
\label{rho_av}
\eea
and corresponds to the average of the indicator function $\rho_i$ over the distribution  $\tilde{P}(\bm{x})$, i.e.
\bea
\hat{\rho}_i=\sum_{\bm{x}}\tilde{P}(\bm{x}) \rho_i
\eea 
Finally the expected size of the MCGC $\hat{R}$ can be calculated directly as
\bea
\hat{R}=\sum_{i=1}^N \hat{\rho}_i.
\eea
Therefore by monitoring $\hat{R}$ versus $p$ it is possible to characterize the typical response of the duplex network to random damage of the nodes
\subsection{Random network ensemble and random realization of the initial damage}

In the previous paragraph we have considered the message passing algorithm that allows to study percolation the mean-field theory of percolation on single instances of duplex networks when the initial damage configuration is  unknown but drawn from the distribution $\tilde{P}(\bm{x})$ given by Eq. $(\ref{random})$.
Here we consider the case in which also the duplex network is unknown but drawn from ensemble of duplex networks formed by two layers each one formed by an uncorrelated network  with degree distribution ${P}(k)$.

In this case we cannot any more consider a message passing algorithm, rather we should consider the mean-field equation \cite{Havlin,Son,Baxter,bianconi2018multilayer}  for percolation of interdependent duplex networks without link overlap obtained by averaging Eqs. ($\ref{mess_av}$) and ($\ref{rho_av}$) over the considered duplex network ensemble.
We indicate with $R$ the probability that a node is in the MCGC. This quantity is given by the average of $\hat{\rho}_i$ over the considered ensemble of duplex networks. Similarly we indicate  with $S^{\prime}$ the probability  that by following a link in one layer we reach a node in the MCGC. This quantity is given by the average of $\hat{\sigma}_{i\to j}$ over the duplex network ensemble. 
Using this notation the  mean field equations for percolation of independent networks without link overlap \cite{Havlin,Son,bianconi2018multilayer} read
\bea
R&=&p\left(1-G_0(1-S^{\prime})\right)^2,\nonumber \\
S^{\prime}&=&p\left(1-G_1(1-S^{\prime})\right)\left(1-G_0(1-S^{\prime})\right),
\label{MF}
\eea
where $G_0(x)$ and $G_1(x)$ are the generating functions 
\bea
G_0(x)&=&\sum_k {P}(k) x^k, \nonumber \\
G_1(x)&=&\sum_k\frac{k{P}(k)}{\avg{k}}x^{k-1}.
\eea
These equations yield a discontinuous phase transition at the percolation threshold $p=p_c$  determined \cite{Havlin,Son,Baxter,bianconi2018multilayer} by the equations $(\ref{MF})$ together with the equation 
\bea
1&=&p\left(1-G_0(1-S^{\prime})\right)\sum_k\frac{k(k-1){P}(k)}{\avg{k}}(1-S^{\prime})^{k-2}\nonumber \\&&+p\left(1-G_1(1-S^{\prime})\right)\sum_k k {P}(k) (1-S^{\prime})^{k-1}.
\eea  
\section{Large deviation theory  of percolation}
\subsection{The large deviation approach to percolation}
While in infinite networks the percolation process is self-averaging, i.e. the typical behaviour characterizes a set of measure one of random instances of the initial damage,  in finite networks deviations from the typical behaviour can be observed. It is therefore necessary to establish the large deviation theory of percolation.
Here our aim is to generalize the framework  proposed in Ref. \cite{rare} for characterizing the large deviation in single networks  in order to  treat the large deviation of percolation in  interdependent multiplex networks.
The large deviation theory of percolation of multiplex networks aims at characterizing  the probability  distribution $\pi(R)$ of size $R$ of the MCGC when the initial damage configuration $\bm{x}$ is chosen with probability $\tilde{P}(\bm{x})$, i.e.
\bea
\pi(R)=\sum_{\bm{x}}\tilde{P}(\bm{x}) \delta({\mathcal R},R),
\label{pi_def}
\eea
where with $\delta(c,d)$ we indicate the  Kronecker delta.
For large network sizes $N\gg 1$ and given value of $p$ the probability $\pi(R)$ has a scaling with the network size $N$ determined by the {\em rate function} $I(R)\geq 0$. In particular we observe the scaling \cite{Touchette}
\bea
\pi({R})\sim e^{-NI({R})}.
\eea
This expression implies that  for  $N\to \infty$ the percolation is self-averaging and $\pi(R)$ is non-zero only for s $R=\hat{R}$ for which $I(R)$ takes its minimum value $I(\hat{R})$.Therefore in the infinite network limit all realizations of the initial damage yield almost surely the same size of the MCGC.
However the rate function $I(R)$ captures the fluctuations in the size of the MCGC that can be observed in finite multiplex networks.
In order to find $I(R)$ we adopt a canonical  approach and we  introduce the partition function $Z=Z(\omega)$ given by 
\bea
Z&=&\sum_{\bm{x}}\tilde{P}(\bm{x}) e^{-\omega {\mathcal R}}.
\eea
Using the definition of $\pi(R)$ given by Eq. $(\ref{pi_def})$ it can be easily shown that 
$Z$ is the generating function of $\pi(R)$ as $Z$  can  be written as 
\bea
Z=\sum_{R}\pi({R})e^{-\omega {R}}.
\label{Zdef}
\eea
The   corresponding free-energy $F$ and the free energy density  $f$ can be calculated as  
\bea
\omega F=\omega Nf=-\log(Z).
\label{F}
\eea
  The  Legendre-Fenchel transform of  the rate function $I(R)$ \cite{Touchette} can be expressed in terms of the free energy density as $\omega f(\omega)$ and we have
\bea
\omega f(\omega)&=&\inf_{R}\left[I(R)+\omega \frac{R}{N}\right].
\label{LF}
\eea
Additionally as long $\omega f(\omega)$ is differentiable,  the Legendre-Fenchel transform $\hat{I}(R)$  of   $\omega f(\omega)$ given by 
\bea
\hat{I}(R)=\sup_{\omega} \left[\omega f(\omega)-\omega \frac{R}{N}\right].
\eea
 fully determines the rate function $I(R)$, i.e. $I(R)=\hat{I}(R)$. 

However when $I(R)$ is non-convex  $\omega f(\omega)$ is not differentiable and the Legendre-Fenchel transform of $\omega f(\omega)$ given by $\hat{I}(R)$ only provides the convex envelop of the rate function $I(R)$ \cite{Touchette}.

Alternatevely it is possible to proceed as proposed in Ref. \cite{Hartmann1,Hartmann_Mezard,Hartmann2} and directly extract the distribution $\pi(R)$ from the biased distribution $\tilde{\pi}(R)$ given by 
\bea
\tilde{\pi}(R)=\frac{1}{Z}\sum_{\bm{x}}\tilde{P}(\bm{x}) e^{-\omega {\mathcal R}}\delta({\mathcal R},R)=\frac{\pi(R)e^{-\omega R}}{Z}
\eea
getting 
\bea
\pi(R)=Ze^{\omega R}\tilde{\pi}(R).
\eea
Therefore the knowledge of the biased distribution $\tilde{\pi}(R)$ which can be sampled with the Markov Chain Monte Carlo method is in principle sufficient to reconstruct the full distribution $\pi(R)$. However this approach is entirely numerical.\\
In the following we will consider the first approach and we will evaluate the partition function $Z$ using the Belief Propagation algorithm.
 
\subsection{The Gibbs measure over messages}

In order to calculate the partition function $Z$ we follow the approach proposed in Ref. \cite{rare} and we consider a Gibbs measure  $P(\bm{\sigma})$ over the set $\bm{\sigma}$ of all messages. The probability distribution $P(\bm{\sigma})$  weights the configuration of the  messages $\bm{\sigma}$ according to  the probability of the corresponding initial damage configurations. Moreover    we introduce a  Lagrangian multiplier $\omega$  conjugated to the size of the MCGC ${\mathcal R}$ that is able to tune the relative weight of the configurations of the messages  corresponding to different sizes of the MCGC.
Therefore the Gibbs measure $P(\bm{\sigma})$ is given by 
\bea
P({\bm{\sigma}})=\frac{1}{Z}\sum_{{\bm{x}}}e^{-\omega {\mathcal R}}\tilde{P}({\bf x})\chi(\bm{\sigma},\bm{x}),
\label{Pss}
\eea
where the function $\chi(\bm{\sigma},\bm{x})$ enforces the message passing Eqs. $(\ref{mes0})$ , i.e.
\bea
\hspace*{-25mm}\chi(\bm{\sigma},\bm{x})=\prod_{\alpha=1,2}\prod_{i=1}^N\prod_{j\in N_{\alpha}(i)}\delta\left(\sigma_{i\to j}^{\alpha},x_i\left[1-\prod_{\ell\in N_{\alpha}(i)\setminus j}(1-\sigma_{\ell\to i}^{\alpha})\right]\left[1-\prod_{\ell\in N_{\beta}(i)}(1-\sigma_{\ell\to i}^{\beta})\right]\right).\nonumber
\eea

The partition function $Z$ in Eq. $(\ref{Pss})$ clearly reduces to $Z$ defined in Eq. (\ref{Zdef}). In fact we have
\bea
Z=\sum_{\bm{\sigma}}\sum_{{\bm{x}}}e^{-\omega {\mathcal R}}\tilde{P}({\bm{x}})\chi({\bm{x}},\bm{\sigma})=\sum_{R}\pi(R)e^{-\omega {R}}.
\eea 
Therefore characterizing the partition function $Z$ and the free energy $F$ (given by Eq. (\ref{F})) corresponding to  the Gibbs measure defined in Eq. $(\ref{Pss})$ allows us to directly calculate the Legendre-Fenchel transform of the rate function $I(R)$.

We note that the Gibbs measure $P(\bm{\sigma})$ can be interpreted as a canonical ensemble, where $\omega$ plays the role of the inverse temperature and ${\mathcal R}$ plays the role of the energy.
Since ${\mathcal R}$ is the sum of the node variable $\rho_i$ and $\rho_i$ can only take two values ($\rho_i=0$- the node does  not belong to the MCGC or $\rho_i=1$ the node does belong to the MCGC) the Gibbs measure  can be interpreted as a a statistical mechanics problem of a  two level system.
It follows that in this case we can investigate the properties of the Gibbs measure for values of $\omega$ that can be also negative. 

For $\omega<0$, the Gibbs measure weights more the {\em buffering } configurations of the  initial damage resulting in a MCGC larger than the typical one.
On the contrary for $\omega>0$ the Gibbs measure weights more the {\em aggravating} configurations of the initial damage resulting in a MCGC smaller than the typical one.
For $\omega=0$ we recover the typical scenario.

The Gibbs  measure $P(\bm{\sigma})$ given by Eq. $(\ref{Pss})$  can also be expressed as
\bea
P(\bm{\sigma})=\frac{1}{Z}\prod_{i=1}^N \psi_i(\bm{\sigma}_i,\omega),
\label{GM}
\eea
where the set of constraints $\psi_i(\bm{\sigma}_i,\omega)$ for $i=1, 2, \ldots, N$ defined over all the messages $\bm{\sigma}_i$ starting or ending to node $i$ read
\bea
\psi_i(\bm{\sigma}_i)&=&(1-p)\prod_{\alpha=1,2}\prod_{j\in N_{\alpha}(i)}\delta(\sigma_{i\to j}^{\alpha},0)\nonumber \\
&&\hspace*{-35mm}+p e^{-\omega \tilde{\rho}_i}\prod_{\alpha=1,2}\prod_{j\in N_{\alpha}(i)}\delta\left(\sigma_{i\to j}^{\alpha},\left[1-\prod_{\ell\in N_{\alpha}(i)\setminus j}(1-\sigma_{\ell \to i}^{\alpha})\right]\left[1-\prod_{\ell\in N_{\beta}(i)\setminus j}(1-\sigma_{\ell \to i}^{\beta})\right]\right).\nonumber
\eea
Here $\delta(m,n)$ indicates the Kronecker delta and $\tilde{\rho}_i$ is given by 
\bea
\tilde{\rho}_i=\prod_{\alpha=1,2}\left[1-\prod_{j\in N_{\alpha}(i)}(1-\sigma_{j\to i}^{\alpha})\right].
\eea
Finally, using  Eq. $(\ref{GM})$ it can be easily shown that  the partition function  $Z$ can be also written as  
\bea
Z=\sum_{\bm{\sigma}}\prod_{i=1}^N \psi_i(\bm{\sigma}_i,\omega).
\eea 

In the following sections we will characterize the large deviation properties of percolation on multiplex networks by calculating the Gibbs measure using Belief Propagation (BP).

\section{Belief Propagation approach}
\subsection{The Belief Propagation equations }

On a locally tree-like duplex network without link overlap the Gibbs distribution $P(\bm{\sigma})$   can be expressed explicitly  using the  Belief Propagation  (BP) algorithm \cite{Mezard_Montanari} by finding the messages 
$\hat{P}_{i\to j}^{\alpha}(\sigma_{i \to j}^{\alpha},\sigma_{j\to i}^{\alpha})$ that each  node $i$  sends to the generic neighbour node  $j$ in layer $\alpha$.
These messages satisfy the following recursive BP equations (see Appendix A for their explicit expression)
\bea
\hspace*{-12mm}\hat{P}_{i\to j}^{\alpha}(\sigma_{i \to j}^{\alpha},\sigma_{j\to i}^{\alpha})=\frac{1}{{\mathcal D}_{i\to j}^{[\alpha]}}\sum_{\bm{\sigma}_i} \psi_i(\bm{\sigma}_i) \prod_{\ell\in N_{\alpha}(i)\setminus j}\hat{P}_{\ell\to i}^{\alpha}(\sigma_{\ell\to i}^{\alpha},\sigma_{i\to \ell}^{\alpha})\prod_{\ell\in N_{\beta}(i)}\hat{P}_{\ell\to i}^{\beta}(\sigma_{\ell\to i}^{\beta},\sigma_{i\to \ell}^{\beta}),\nonumber
\label{BP}
\eea
where $\alpha\neq \beta$ and where ${\mathcal D}_{i\to j}^{[\alpha]}$ are normalization constants enforcing the normalization condition 
\bea
\sum_{\sigma_{i\to j}^{\alpha}=0,1}\sum_{\sigma_{j\to i}^{\alpha}=0,1}\hat{P}_{i\to j}^{\alpha}(\sigma_{i\to j}^{\alpha},\sigma_{j\to i}^{\alpha})=1.
\label{normBP}
\eea
In the Bethe approximation, valid on locally tree-like networks the probability distribution $P(\bm{\sigma})$ is given by \cite{Mezard_Montanari}
\bea
P(\bm{\sigma})&=&\prod_{i=1}^N {\mathcal P}_i(\bm{\sigma}_i)\left(\prod_{\alpha=1,2}\prod_{<i,j>_{\alpha}}{\mathcal P}_{ij}^{\alpha}(\sigma_{i \to j}^{\alpha},\sigma_{j \to i}^{\alpha})\right)^{-1}
\label{uno}
\eea
where  ${\mathcal P}_{i}(\bm{\sigma}_i)$ and ${\mathcal P}_{ij}^{\alpha}(\sigma_{i \to j}^{\alpha},\sigma_{j \to i}^{\alpha})$ indicate the marginal distribution of nodes and links and are given  by \cite{Mezard_Montanari}
 \bea
&&{\mathcal P}_{ij}^{\alpha}(\sigma_{i \to j}^{\alpha},\sigma_{j \to i}^{\alpha})=\frac{1}{\mathcal{C}_{ij}^{\alpha}}\hat{P}_{i\to j}^{\alpha}(\sigma_{i \to j}^{\alpha},\sigma_{j \to i}^{\alpha})\hat{P}_{j\to i}^{\alpha}(\sigma_{j \to i}^{\alpha},\sigma_{i \to j}^{\alpha}),\nonumber \\
&& {\mathcal P}_{i}(\bm{\sigma}_i)=\frac{1}{\mathcal{C}_{i}} \psi_i(\bm{\sigma}_i)\prod_{\alpha=1,2}\prod_{j\in N_{\alpha}(i)}\hat{P}_{j\to i}^{\alpha}(\sigma_{j \to i},\sigma_{i \to j}),
\label{marginals}
\eea
with ${\mathcal C}_i$ and ${\mathcal C}_{ij}^{[\alpha]}$ indicating normalization constants (see Appendix B for their explicit expression).

\subsection{Free energy}

The free energy $F$  given by Eq. (\ref{F})
 can be found by minimizing the Gibbs free energy $F_{Gibbs}$  given by 
\bea
\omega F_{Gibbs}=\sum_{\bm{\sigma}}P(\bm{\sigma})\ln\left(\frac{P(\bm{\sigma})}{\psi(\bm{\sigma})}\right),
\eea
 where $\psi(\bm{\sigma})$  indicates the set of constraints
\bea
\psi(\bm{\sigma})=\prod_{i=1}^N\psi_i(\bm{\sigma}_i).
\eea
In fact it can be easily shown that the   Gibbs free energy $F_{Gibbs}$ is minimal when calculated over  the probability distribution $P(\bm{\sigma})$ given by Eq. $(\ref{GM})$ and that its minimum value is  
\bea
\omega F_{Gibbs}=\omega F=-\ln Z.
\eea
On a locally tree-like duplex network the Gibbs measure $P(\bm{\sigma})$ reduces to Eq. $(\ref{uno})$, and it can be easily  see that the free energy can be expressed as 
\begin{equation}\label{betheF}
\omega F=\sum_{\alpha=1,2}\sum_{<i,j>_{\alpha}}\log \left({\mathcal C}_{ij}^{\alpha}\right) - \sum_{i=1}^N \log({\mathcal C}_i).
\end{equation}
Given the explicit expression of ${\mathcal C}_{ij}^{[\alpha]}$ and of ${\mathcal C}_i $   in terms of the messages (see Appendix B), the free energy $F$ can be calculated easily when the BP equations have been solved.

\subsection{Energy and Specific Heat}

The energy and the specific heat corresponding to the Gibbs measure $P(\bm{\sigma})$ have also a very important interpretation in terms of the underlying percolation process.
The energy is given by the  the average size of the MCGC $R$.  In fact we have 
\bea
R&=&\sum_{\bm{\sigma}}{\mathcal R}P(\bm{\sigma})=-\frac{\partial \ln Z}{\partial \omega}.
\eea
The mean-field percolation transition correspond to the phase transition from a non-percolating phase with $R=0$ to a percolating phase $R>0$ at $p=p_c$ and $\omega=0$.
However the present approach allows to consider the full line of critical points $(p^{\star},\omega^{\star})$ at which the transition is observed when the large deviations are considered.

Following a statistical mechanics definition, we can also define the    specific heat $C$   as 
\bea
\frac{C}{\omega^2}&=&-\frac{\partial R}{\partial \omega}. \nonumber \\
\eea
The specific heat has the immediate interpretation  in terms of  the variance in the   size of  the MCGC, i.e.
\bea
\frac{C}{\omega^2}=\left(\sum_{\bm{\sigma}}{\mathcal R}^2P(\bm{\sigma})\right)-\left(\sum_{\bm{\sigma}}{\mathcal R}P(\bm{\sigma})\right)^2.\nonumber
\eea
Both $R$ and $C/\omega^2$ can be derived from the message passing algorithm. Indeed we have 
\bea
R&=&\sum_i r_i,\label{R} \\
\frac{C}{\omega^2}&=&\sum_{i=1}^N r_i\left(1-r_i\right)
\label{C}
\eea
where 
\bea
r_i&=&\sum_{\bm{\sigma}}\rho_i P(\bm{\sigma})
\eea
indicating the probability that node $i$ is in the giant component
 is given by 
 \bea
 r_i=\frac{z_i}{{\mathcal C}_i},
 \eea
where the explicit expression of $z_i$ and ${\mathcal C}_i$ is given in Appendix B.
The quantity $C/\omega^2$  can be also interpreted as the expected fraction of nodes that given  two random realizations of the initial damage are found  in the MCGC  in one realization but not in the other.
This quantity generalizes the measure proposed in Ref. \cite{Fluctuations} to characterize the fluctuations in percolation in single networks.
\subsection{The typical scenario}

It is instructive to see that the proposed large deviation theory of percolation reduces to the mean-field theory of percolation in the typical scenario obtained by putting $\omega=0$. In this case we obtain that the BP equations have solution with
\bea
\hat{P}_{i\to j}^{\alpha}(0,0)=\hat{P}_{i\to j}^{\alpha}(0,1),\nonumber \\
\hat{P}_{i\to j}^{\alpha}(1,1)=\hat{P}_{i\to j}^{\alpha}(1,0),\nonumber \\
\eea
and the BP equations for $\omega=0$ reduces to Eqs.(\ref{mess_av}) when we put
\bea
\hat{\sigma}_{i\to j}^{\alpha}=\hat{P}_{i\to j}^{\alpha}(1,1)+\hat{P}_{i\to j}^{\alpha}(1,0).
\eea
Similarly it is easy to show that $r_i$ reduces to $\hat{\rho}_i$ given by Eq. ($\ref{rho_av}$).

\section{Numerical results}

\begin{figure} 
    \includegraphics[width=0.90\columnwidth]{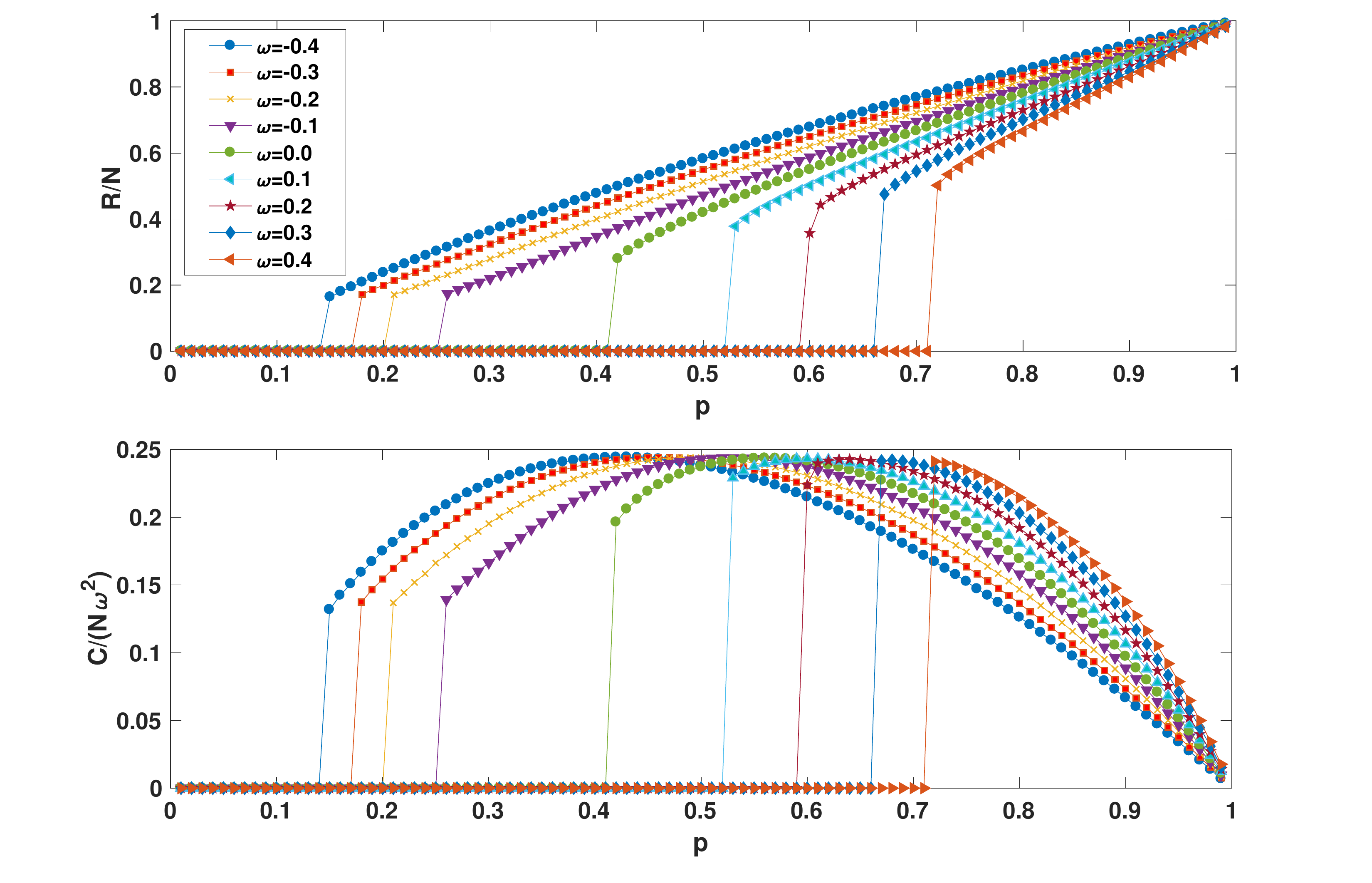}
	\caption{The average fraction of nodes of the MCGC $R/N$ and the normalized fluctuations $C/(N\omega^2)$ of the size of the MCGC are plotted as a function of $p$ for different values of $\omega$. The considered multiplex network is  a duplex Poisson network with average degree $z=6$ and total number of nodes $N=100$.}
	\label{fig:RC}
\end{figure}

Here we consider the results obtained by running the BP algorithm over a Poisson duplex network with average degree $z=6$ and network size $N=100$ (see Figure $\ref{fig:RC}$).
The typical scenario observed for $\omega=0$  gives a discontinuous transition of the average size $R$ of the MCGC at $p=p_c=0.409235\ldots$ with $R=R_c=0.209405\ldots$ for the considered duplex network as predicted by the mean-field theory of percolation.
The fluctuations of the size of the MCGC measured by $C/\omega^2$ for $\omega\to 0$ have also a jump at $p=p_c$. Therefore as we approach the critical percolation threshold from above (i.e. for $p\to p_c^{+}$) we observe significant fluctuations in the size of the MCGC. However these fluctuations are maximal only for larger values of $p$. 

Interestingly the  BP results allow us to investigate also how the average size of the MCGC $R$ and its fluctuations  $C/\omega^2$ change if we deviate from the typical scenario, i.e. for $\omega\neq 0$. 
For $\omega<0$ when we consider buffering configuration of the initial damage we observe a transition for lower values of $p$, for $\omega>0$ we observe a transition for larger value of $p$. In both cases the transition remains discontinuous in the investigated range of values of $\omega$. Additionally we note that for buffering configurations,  as $\omega$ decreases the  discontinuous jump in the size of the MCGC $R $  appears to reach a constant value. This suggests that the discontinuity might be preserved even beyond the observed range of values of $\omega$.
The fluctuations in the size of the MCGC $C/\omega^2$ observed at the transition point  increase for aggravating configuration of the damage. Moreover for aggravating configuration of the damage these fluctuations can achieve their maximum at the transition point itself.

\begin{figure} 
    \includegraphics[width=0.90\columnwidth]{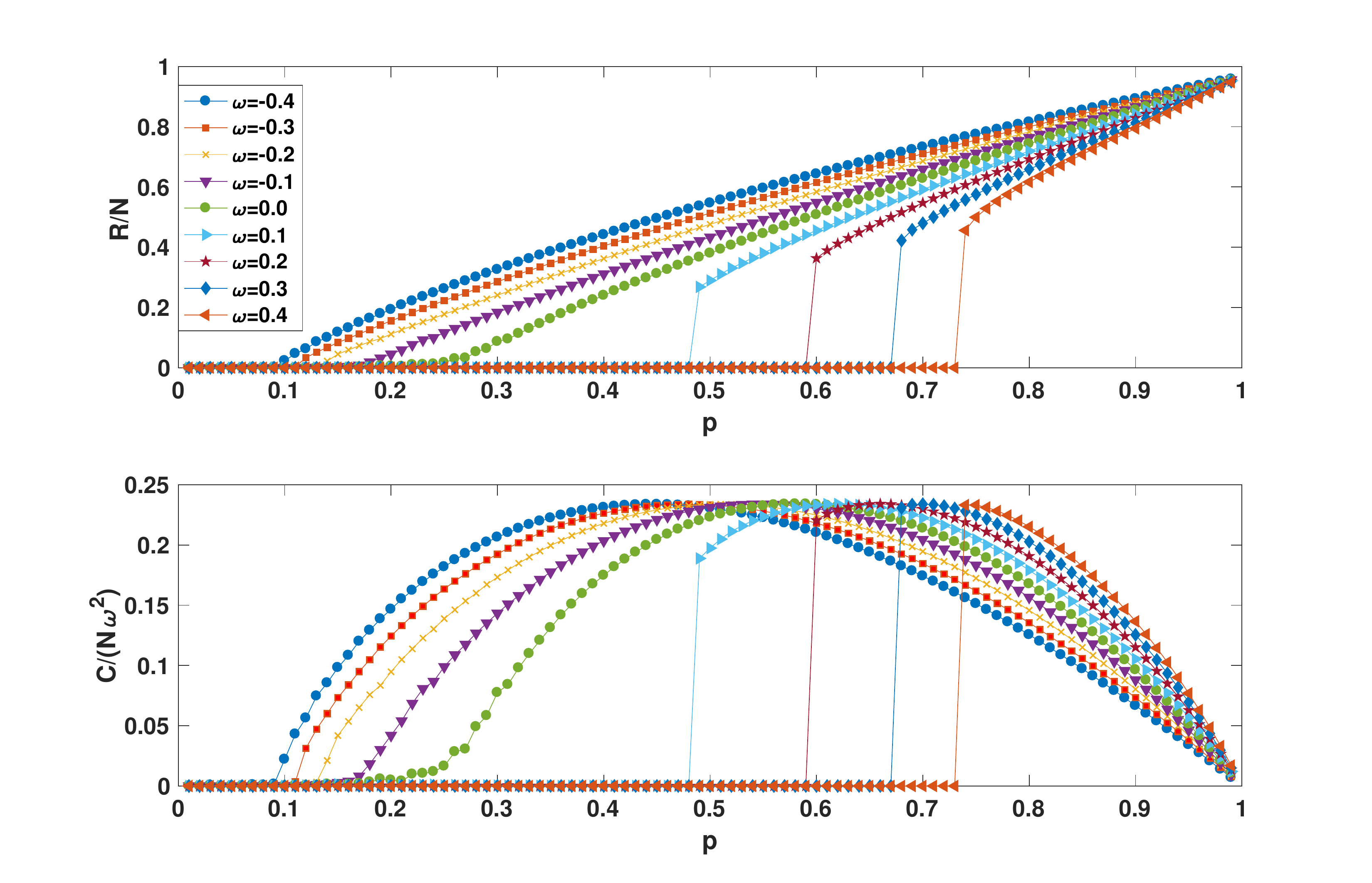}
	\caption{The average fraction of nodes of the giant component $R/N$ and the normalized fluctuations $C/(N\omega^2)$ of the size of the giant component  are plotted as a function of $p$ for different values of $\omega$. The considered network is  a single Poisson network with average degree $z=4$ and total number of nodes $N=100$ with percolation threshold $p_c=1/4$. }
	\label{fig:single}
\end{figure}

 In Figure $\ref{fig:single}$ we compare the results obtained for the duplex networks to the results that can be obtained by studying the large deviations of percolation on single networks as discussed in Ref. \cite{rare}. In the duplex network we observe a discontinuous jump of $C/\omega^2$ also in the typical scenario $\omega=0$ while in single networks $C/\omega^2$ is continuous at the transition point. Therefore the observed discontinuity of $C/\omega^2$ at $p=p_c$ and $\omega=0$ is a purely multiplex network phenomenon not observed in percolation of single networks. In fact the presence of significant fluctuations of the percolation order parameter at the percolation transition can be only observed if the order parameter has a discontinuous jump and in this case only for  $p\to p_c^{+}$ when $R\to R_c>0$. Finally we note that in the multiplex case the discontinuity of $C/\omega^2$ extend also for negative value of $\omega$ corresponding to buffering configurations of the initial damage while in single networks we observe a continuous behaviour of $C/\omega^2$.

\begin{figure} 
    \includegraphics[width=0.90\columnwidth]{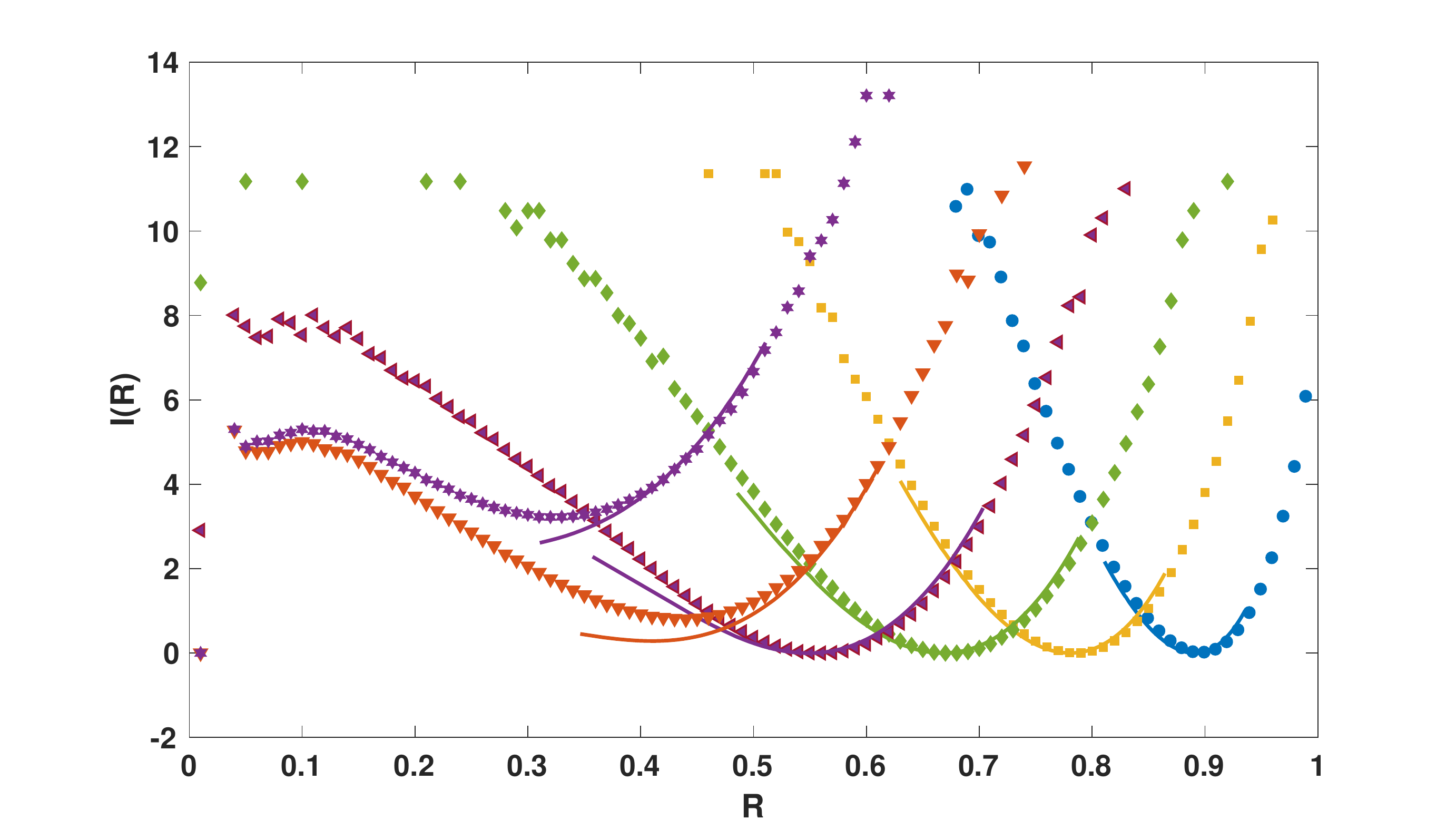}
	\caption{The rate function $I(R)$ measured starting from $10^6$ random realizations of the initial damage (symbols)  and compared to its convex envelop obtained using the BP algorithm (solid lines). The different color and symbols correspond to different values of $p$: $p=0.9$ (blue circles), $p=0.8$ (yellow squares), $p=0.7$ (green diamonds), $p=0.6$ (purple triangles), $p=0.5$ (red triangles pointing left), $p=0.4$ (purple stars). The considered multiplex network is  a duplex Poisson network with average degree $z=6$ and total number of nodes $N=100$.}
	\label{fig:IR}
\end{figure}

In order to test the validity of the BP algorithm, we have compared the rate function $I(R)$ measured starting from $10^6$ initial damage configurations performed on the same duplex network instance on which the BP algorithm is run, with  $\hat{I}(R)$ provided by the BP algorithm finding very good results (see Figure $\ref{fig:IR}$).
We notice that while for large values of $p$ the rate function $I(R)$ is convex and therefore $I(R)=\hat{I}(R)$, as we approach the percolation transition the rate function $I(R)$ becomes non-convex and $\hat{I}(R)$ only provides the convex envelop of $I(R)$.

\section{Conclusions}
In this paper we have characterized the large deviation of percolation of interdependent multiplex networks without link overlap using a Belief Propagation algorithm.
In the typical scenario, well captured by the mean-field theory of percolation we observe a discontinuous percolation transition. Our analysis reveals that  when we depart from the typical scenario the percolation transition can occur for values of $p$ significantly distant from the percolation threshold in the typical scenario while remaining discontinuous both for buffering configurations of the random damage and aggravating ones.
Interestingly our study show a significant difference of interdependent percolation with respect to percolation in single layers. In fact when we consider percolation of isolated networks the fluctuations in the size of the giant component go to zero at the critical point. However in interdependent percolation we can observe significant fluctuations of the size of the MCGC when we approach the critical point from above, i.e. for $p\to p_c^{+}$.
Finally we have compared the rate function $I(R)$ of observing a MCGC of size $R$ with the predicted $\hat{I}(R)$ obtained using the proposed canonical BP algorithm.  As observed in the single layer scenario the rate function $I(R)$ becomes bimodal for small values of $p$, and therefore the proposed canonical BP algorithm can only capture its convex envelop, i.e. we have $\hat{I}(R)\neq I(R)$.

The proposed method can be extended in different ways by considering percolation and directed percolation in multiplex network with link overlap and by developing a microcanonical approach in which instead of introducing the Lagrangian multiplier $\omega$ fixing the size of the MCGC in average, we consider the corresponding hard constraint.

\section*{Acknowledgements}
We thank Alexander Hartmann, Francesco Coghi, Giorgio Parisi, Riccardo Zecchina and Robert Ziff for  interesting discussions.
 
\appendix
\section{Explicit Belief Propagation equations}
In this appendix we provide the explicit expression of the Belief Propagation equations $(\ref{BP})$. When degree $k_i^{[\alpha]}>1$ and degree $k_i^{[\beta]}>0$ these equations  read
\bea
&&\hspace*{-20mm}{\mathcal D}_{i\to j}^{[\alpha]}\hat{P}_{i\to j}^{\alpha}(0,0)=(1-p)\left\{\prod_{\ell \in N_{\alpha}(i)\setminus j}\left[\hat{P}_{\ell \to i}^{\alpha}(0,0)+\hat{P}_{\ell\to i }^{\alpha}(1,0)\right]\right\}\left\{\prod_{\ell \in N_{\beta}(i)}\left[\hat{P}_{\ell \to i}^{\beta}(0,0)+\hat{P}_{\ell\to i }^{\beta}(1,0)\right]\right\}\nonumber \\
&&\hspace*{-20mm}+p\left\{\prod_{\ell \in N_{\alpha}(i)\setminus j}\left[\hat{P}_{\ell \to i}^{\alpha}(0,0)+\hat{P}_{\ell\to i }^{\alpha}(1,0)\right]\right\}\left\{\prod_{\ell \in N_{\beta}(i)}\hat{P}_{\ell \to i}^{\beta}(0,0)\right\}\nonumber \\
&&\hspace*{-20mm} +p\left\{\prod_{\ell \in N_{\beta}(i)}\left[\hat{P}_{\ell \to i}^{\beta}(0,0)+\hat{P}_{\ell\to i }^{\beta}(1,0)\right]-\prod_{\ell \in N_{\beta}(i)}\hat{P}_{\ell \to i}^{\beta}(0,0)\right\}\left\{\prod_{\ell \in N_{\alpha}(i)\setminus j}\hat{P}_{\ell \to i}^{\alpha}(0,0)\right\},\nonumber \\
&&\hspace*{-20mm}{\mathcal D}_{i\to j}^{[\alpha]} \hat{P}_{i\to j}^{\alpha}(0,1)=(1-p)\left\{\prod_{\ell \in N_{\alpha}(i)\setminus j}\left[\hat{P}_{\ell \to i}^{\alpha}(0,0)+\hat{P}_{\ell\to i }^{\alpha}(1,0)\right]\right\}\left\{\prod_{\ell \in N_{\beta}(i)}\left[\hat{P}_{\ell \to i}^{\beta}(0,0)+\hat{P}_{\ell\to i }^{\beta}(1,0)\right]\right\}\nonumber \\
&& \hspace*{-20mm} +p\left\{\prod_{\ell \in N_{\alpha}(i)\setminus j}\left[\hat{P}_{\ell \to i}^{\alpha}(0,0)+\hat{P}_{\ell\to i }^{\alpha}(1,0)\right]\right\}\left\{\prod_{\ell \in N_{\beta}(i)}\hat{P}_{\ell \to i}^{\beta}(0,0)\right\}\nonumber \\
&&\hspace*{-20mm} +pe^{-\omega}\left\{\prod_{\ell \in N_{\beta}(i)}\left[\hat{P}_{\ell \to i}^{\beta}(0,1)+\hat{P}_{\ell\to i }^{\beta}(1,1)\right]-\prod_{\ell \in N_{\beta}(i)}\hat{P}_{\ell \to i}^{\beta}(0,1)\right.\nonumber \\
&&\hspace*{-20mm} \left.+\sum_{\ell \in N_{\beta}(i)}\left[\hat{P}_{\ell\to i}^{\beta}(1,0)-\hat{P}_{\ell \to i}^{\beta}(1,1)\right]\prod_{\ell'\in N_{\beta}(i)\setminus \ell}\hat{P}_{\ell'\to i}^{\beta}(0,1)\right\}\nonumber \\
&&\hspace*{-20mm} \left\{\prod_{\ell \in N_{\alpha}(i)\setminus j}\hat{P}_{\ell \to i}^{\alpha}(0,1)\right\},\nonumber \\
&&\hspace*{-20mm} {\mathcal D}_{i\to j}^{[\alpha]} \hat{P}_{i\to j}^{\alpha}(1,0)=pe^{-\omega}\left\{\prod_{\ell \in N_{\beta}(i)}\left[\hat{P}_{\ell \to i}^{\beta}(0,1)+\hat{P}_{\ell\to i }^{\beta}(1,1)\right]-\prod_{\ell \in N_{\beta}(i)}\hat{P}_{\ell \to i}^{\beta}(0,1)\right.\nonumber \\
&&\hspace*{-20mm} +\left.\sum_{\ell \in N_{\beta}(i)}\left[\hat{P}_{\ell\to i}^{\beta}(1,0)-\hat{P}_{\ell \to i}^{\beta}(1,1)\right]\prod_{\ell'\in N_{\beta}(i)\setminus \ell}\hat{P}_{\ell'\to i}^{\beta}(0,1)\right\}\nonumber \\
 &&\hspace*{-20mm} \times\left\{\prod_{\ell \in N_{\alpha}(i)\setminus j}\left[\hat{P}_{\ell \to i}^{\alpha}(0,1)+\hat{P}_{\ell\to i }^{\alpha}(1,1)\right]-\prod_{\ell \in N_{\alpha}(i)\setminus j}\hat{P}_{\ell \to i}^{\alpha}(0,1)\right.\nonumber \\
 &&\hspace*{-20mm} \left.+\sum_{\ell \in N_{\alpha}(i)\setminus j}\left[\hat{P}_{\ell\to i}^{\alpha}(1,0)-\hat{P}_{\ell \to i}^{\alpha}(1,1)\right]\prod_{\ell'\in N_{\alpha}(i)\setminus  j,\ell}\hat{P}_{\ell'\to i}^{\alpha}(0,1)\right\}\nonumber \\
&&\hspace*{-20mm} {\mathcal D}_{i\to j}^{[\alpha]}\hat{P}_{i \to j}^{\alpha}(1,1)=pe^{-\omega}\left\{\prod_{\ell \in N_{\beta}(i)}\left[\hat{P}_{\ell \to i}^{\beta}(0,1)+\hat{P}_{\ell\to i }^{\beta}(1,1)\right]-\prod_{\ell \in N_{\beta}(i)}\hat{P}_{\ell \to i}^{\beta}(0,1)\right.\nonumber \\
&&\hspace*{-20mm} \left.+\sum_{\ell \in N_{\beta}(i)}\left[\hat{P}_{\ell\to i}^{\beta}(1,0)-\hat{P}_{\ell \to i}^{\beta}(1,1)\right]\prod_{\ell'\in N_{\beta}(i)\setminus \ell}\hat{P}_{\ell'\to i}^{\beta}(0,1)\right\}\nonumber \\
 &&\hspace*{-20mm} \times\left\{\prod_{\ell \in N_{\alpha}(i)\setminus j}\left[\hat{P}_{\ell \to i}^{\alpha}(0,1)+\hat{P}_{\ell\to i }^{\alpha}(1,1)\right]-\prod_{\ell \in N_{\alpha}(i)\setminus j}\hat{P}_{\ell \to i}^{\alpha}(0,1)\right\}.
\eea
Here ${\mathcal D}_{i\to j}^{[\alpha]}$ are normalization constants fixed by the conditions expressed in Eq. ($\ref{normBP}$).
When degree $k_i^{[\alpha]}=1$ and degree $k_i^{[\beta]}>0$ these equations read
\bea
&&\hspace*{-20mm}{\mathcal D}_{i\to j}^{[\alpha]}\hat{P}_{i\to j}^{\alpha}(0,0)=\left\{\prod_{\ell \in N_{\beta}(i)}\left[\hat{P}_{\ell \to i}^{\beta}(0,0)+\hat{P}_{\ell\to i }^{\beta}(1,0)\right]\right\},\nonumber \\
&&\hspace*{-20mm}{\mathcal D}_{i\to j}^{[\alpha]} \hat{P}_{i\to j}^{\alpha}(0,1)=(1-p)\left\{\prod_{\ell \in N_{\beta}(i)}\left[\hat{P}_{\ell \to i}^{\beta}(0,0)+\hat{P}_{\ell\to i }^{\beta}(1,0)\right]\right\} +p\left\{\prod_{\ell \in N_{\beta}(i)}\hat{P}_{\ell \to i}^{\beta}(0,0)\right\}\nonumber \\
&&\hspace*{-20mm} +pe^{-\omega}\left\{\prod_{\ell \in N_{\beta}(i)}\left[\hat{P}_{\ell \to i}^{\beta}(0,1)+\hat{P}_{\ell\to i }^{\beta}(1,1)\right]-\prod_{\ell \in N_{\beta}(i)}\hat{P}_{\ell \to i}^{\beta}(0,1)\right.\nonumber \\
&&\hspace*{-20mm} \left.+\sum_{\ell \in N_{\beta}(i)}\left[\hat{P}_{\ell\to i}^{\beta}(1,0)-\hat{P}_{\ell \to i}^{\beta}(1,1)\right]\prod_{\ell'\in N_{\beta}(i)\setminus \ell}\hat{P}_{\ell'\to i}^{\beta}(0,1)\right\},\nonumber \\
&&\hspace*{-20mm}  \hat{P}_{i\to j}^{\alpha}(1,0)=0,\nonumber \\
&&\hspace*{-20mm} \hat{P}_{i \to j}^{\alpha}(1,1)=0.
\eea

When for arbitrary degree $k_i^{[\alpha]}\geq 1$ and degree $k_i^{[\beta]}=0$ these equations read
\bea
&&\hspace*{-20mm}  \hat{P}_{i\to j}^{\alpha}(0,0)=\frac{1}{2},\nonumber \\
&&\hspace*{-20mm} \hat{P}_{i \to j}^{\alpha}(0,1)=\frac{1}{2},\nonumber \\
&&\hspace*{-20mm}  \hat{P}_{i\to j}^{\alpha}(1,0)=0,\nonumber \\
&&\hspace*{-20mm} \hat{P}_{i \to j}^{\alpha}(1,1)=0.
\eea

\section{Explicit expression of ${\mathcal C}_{ij}^{\alpha}$, ${\mathcal C}_i$ and $z_i$}
In this appendix we give the explict expression of ${\mathcal C}_{ij}^{\alpha}$, ${\mathcal C}_i$ and $z_i$ in terms of the messages that solve the BP equations. In particular by normalizing the marginals in Eqs. $(\ref{marginals})$ we obtain
\bea
&&\hspace*{-20mm}{\mathcal C}_{ij}^{\alpha}=\hat{P}^{\alpha}_{i\to j}(0,0)\hat{P}^{\alpha}_{j\to i}(0,0)+\hat{P}^{\alpha}_{i\to j}(0,1)\hat{P}^{\alpha}_{j\to i}(1,0)+\hat{P}^{\alpha}_{i\to j}(1,0)\hat{P}^{\alpha}_{j\to i}(0,1)+\hat{P}^{\alpha}_{i\to j}(1,1)\hat{P}^{\alpha}_{j\to i}(1,1),\nonumber \\
&&\hspace*{-20mm}{\mathcal C}_i=(1-p)\left\{\prod_{\ell \in N_{\alpha}(i)}\left[\hat{P}_{\ell \to i}^{\alpha}(0,0)+\hat{P}_{\ell\to i }^{\alpha}(1,0)\right]\right\}\left\{\prod_{\ell \in N_{\beta}(i)}\left[\hat{P}_{\ell \to i}^{\beta}(0,0)+\hat{P}_{\ell\to i }^{\beta}(1,0)\right]\right\}\nonumber \\
&&\hspace*{-20mm}+p\left\{\prod_{\ell \in N_{\alpha}(i)}\left[\hat{P}_{\ell \to i}^{\alpha}(0,0)+\hat{P}_{\ell\to i }^{\alpha}(1,0)\right]\right\}\left\{\prod_{\ell \in N_{\beta}(i)}\hat{P}_{\ell \to i}^{\beta}(0,0)\right\}\nonumber \\  
&&\hspace*{-20mm}+p\left\{\prod_{\ell \in N_{\beta}(i)}\left[\hat{P}_{\ell \to i}^{\beta}(0,0)+\hat{P}_{\ell\to i }^{\beta}(1,0)\right]-\prod_{\ell \in N_{\beta}(i)}\hat{P}_{\ell \to i}^{\beta}(0,0)\right\}\left\{\prod_{\ell \in N_{\alpha}(i)}\hat{P}_{\ell \to i}^{\alpha}(0,0)\right\}+z_i,\nonumber \\
&&\hspace*{-20mm}z_i=pe^{-\omega}\left\{\prod_{\ell \in N_{\beta}(i)}\left[\hat{P}_{\ell \to i}^{\beta}(0,1)+\hat{P}_{\ell\to i }^{\beta}(1,1)\right]-\prod_{\ell \in N_{\beta}(i)}\hat{P}_{\ell \to i}^{\beta}(0,1)\right.\nonumber \\
&&\hspace*{-20mm}+\left.\sum_{\ell \in N_{\beta}(i)}\left[\hat{P}_{\ell\to i}^{\beta}(1,0)-\hat{P}_{\ell \to i}^{\beta}(1,1)\right]\prod_{\ell'\in N_{\beta}(i)\setminus \ell}\hat{P}_{\ell'\to i}^{\beta}(0,1)\right\}\nonumber \\
 &&\hspace*{-20mm}\times\left\{\prod_{\ell \in N_{\alpha}(i)\setminus j}\left[\hat{P}_{\ell \to i}^{\alpha}(0,1)+\hat{P}_{\ell\to i }^{\alpha}(1,1)\right]-\prod_{\ell \in N_{\alpha}(i)\setminus j}\hat{P}_{\ell \to i}^{\alpha}(0,1)\right.\nonumber \\
 &&\hspace*{-20mm}\left.+\sum_{\ell \in N_{\alpha}(i)}\left[\hat{P}_{\ell\to i}^{\alpha}(1,0)-\hat{P}_{\ell \to i}^{\alpha}(1,1)\right]\prod_{\ell'\in N_{\alpha}(i),\ell}\hat{P}_{\ell'\to i}^{\alpha}(0,1)\right\}
\eea

\end{document}